\documentclass[10pt,twocolumn]{IEEEtran}
\usepackage{setspace}
\usepackage{amsmath,amssymb}    
\usepackage{graphicx}   
\usepackage{verbatim}   
\usepackage{color}      
\usepackage{epstopdf}
\usepackage{float}
\usepackage{cite}
\usepackage{amssymb,amsthm}
\usepackage{graphicx}
\usepackage[font=footnotesize]{subfig}

\newtheorem{prob-statement}{Problem}

\newtheorem{thrm}{Theorem}

\DeclareMathOperator*{\argmin}{arg\,min}

\newcommand*{\Scale}[2][4]{\scalebox{#1}{$#2$}}%

\begin{document}

\title{Strategic Communication Between Prospect Theoretic Agents over a Gaussian Test Channel\thanks{This research was supported in part by the Army Research Office (ARO) under Grant W911NF-16-1-0485, and by National Science Foundation (NSF) under Grant 1619339.}}

\author{
\IEEEauthorblockN{{\large Venkata Sriram Siddhardh Nadendla, Emrah Akyol, Cedric Langbort and Tamer Ba\c{s}ar}}
\\[1ex]
\IEEEauthorblockA{Coordinated Science Laboratory, University of Illinois at Urbana-Champaign,
Urbana, IL \\ Email: \{nadendla, akyol, langbort, basar1\}@illinois.edu}
}

\maketitle

\begin{abstract}
In this paper, we model a Stackelberg game in a simple Gaussian test channel where a human transmitter (leader) communicates a source message to a human receiver (follower). We model human decision making using prospect theory models proposed for continuous decision spaces. Assuming that the value function is the squared distortion at both the transmitter and the receiver, we analyze the effects of the weight functions at both the transmitter and the receiver on optimal communication strategies, namely encoding at the transmitter and decoding at the receiver, in the Stackelberg sense. We show that the optimal strategies for the behavioral agents in the Stackelberg sense are identical to those designed for unbiased agents. At the same time, we also show that the prospect-theoretic distortions at both the transmitter and the receiver are both larger than the expected distortion, thus making behavioral agents less contended than unbiased agents. Consequently, the presence of cognitive biases increases the need for transmission power in order to achieve a given distortion at both transmitter and receiver.
\end{abstract}


\section{Introduction \label{sec: Introduction}}
Communication systems have been actively studied by several research scholars under various application scenarios over the past 7 decades \cite{Shannon1948}. In the field of information theory \cite{Book-Cover} and communication technology \cite{Book-Lee}, researchers have focused on the design and analysis of communication systems under synergistic conditions between the senders and the receivers. Such a cooperative behavior may not always exist in practice, especially in scenarios where information exchange takes place in a competitive scenario. For example, in a typical bargaining setup, the seller does not have the same intention as the buyer to share information regarding a given product-of-interest. In such a case, the seller may even find an incentive to falsify product information before sharing it with the buyer. Communication in the absence of synergy between the sender and the receiver has been extensively studied in the economics literature under the labels of \emph{signaling games} \cite{Spence1973}, \emph{strategic communication}\cite{Crawford1982} and \emph{cheap talk}\cite{Farrell1996}. 

Strategic communication was first studied by Crawford and Sobel in \cite{Crawford1982}, where they have considered a non-synergistic setting due to non-identical utilities at the sender and the receiver. More specifically, such a disparity arises in the presence of some private information at the sender, which in turn influences the sender's utility. On the other hand, the receiver has a different utility, which is independent of sender's private information. Crawford and Sobel have found that both the sender and the receiver employ quantization strategies at Nash equilibrium. In contrast, Kamenica and Gentzkow have studied the problem of Bayesian persuasion in \cite{Kamenica2011}, where they have derived necessary and sufficient conditions for the existence of a transmitted signal that benefits the sender. Recently, Akyol \emph{et al.} have adapted traditional settings such as compression, coded and uncoded communication studied by the information theory community, into a strategic setting in \cite{Akyol2017b}. Furthermore, Sarita\c{s} \emph{et al.} have extended Crawford-Sobel's framework to multidimensional sources in the presence of a Gaussian test channel in \cite{Saritas2017}, where the authors have investigated conditions for existence/uniqueness of both Nash and Stackelberg equilibria. Unlike Crawford and Sobel, \cite{Kamenica2011}, \cite{Akyol2017} and \cite{Saritas2017} have studied Stackelberg games in strategic communication, where the receiver (follower) optimizes its decoding strategy based on the encoding policy employed by the sender (leader). Particularly, in the context of a simple communication setting over Gaussian test channels, they have showed that the optimal strategies in the strategic sense are zero-delay linear mappings (uncoded transmissions), which are also optimal in the Shannon sense. 

In the context of decentralized control systems, a different kind of misalignment has been investigated in the context of team decision problems. Ba\c{s}ar had studied a multiplayer problem in the presence of mismatched beliefs (prior distributions) regarding an unknown source and individual (unshared) measurements in \cite{Basar1985}, and had shown that the optimal strategies in the sense of Nash equilibrium are linear mappings, while they are generally nonlinear mappings in the sense of Stackelberg equilibria in the presence of Gaussian priors. Recently, Akyol \emph{et al.} studied a Bayesian manipulation framework in \cite{Akyol2017}, where the motives of the sender and the receiver are misaligned due to two reasons: \emph{mismatched beliefs} and the presence of \emph{private information} at the sender. Although the optimal strategies in the Stackelberg sense are still linear mappings, the sender can take additional leverage from the mismatched priors and send false information to the receiver. Such a strategy improves the sender's utility, while deteriorating the utility at the receiver.

In this paper, we consider a different kind of misalignment between the sender and the receiver, which arises due to the differences in their behavioral psychology. We consider a simple Gaussian test channel where both the transmitter and the receiver are human agents, whose utilities are modeled using prospect theory. The main difference between the mismatch models studied in the past and our current model is that the cognitive biases of the behavioral agents also distort the channel's transition probability distributions, along with the source distributions. In this paper, we focus on the nonlinear reaction to probabilities by human agents, which is well-captured by weighting functions in prospect theory models \cite{Karmarkar1978, Kahneman1979, Prelec1998, Rieger2008}. We show that the transmitter can no longer take leverage from the behavioral mismatch, since the optimal strategies at the transmitter and the receiver are identical to that of an unbiased agent. At the same time, we also show that this behavioral mismatch has a positive impact on the agent's distortion functions, making them more satisfied than the case of an unbiased agent.

\begin{figure*}[!t]
	\centerline
    {
    	\subfloat[Typical weight-function]{\includegraphics[width=3.1in]{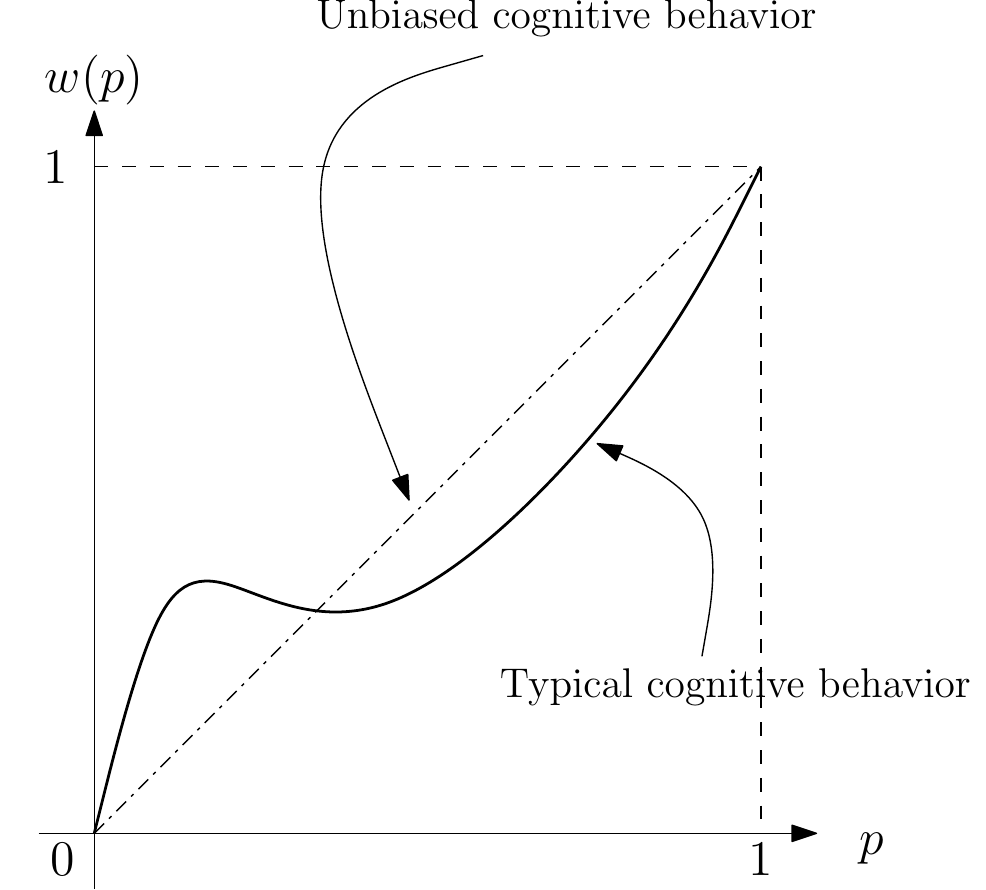}%
        \label{Fig: weight}}
        \hfil
        \subfloat[Typical value-function]{\includegraphics[width=3.3in]{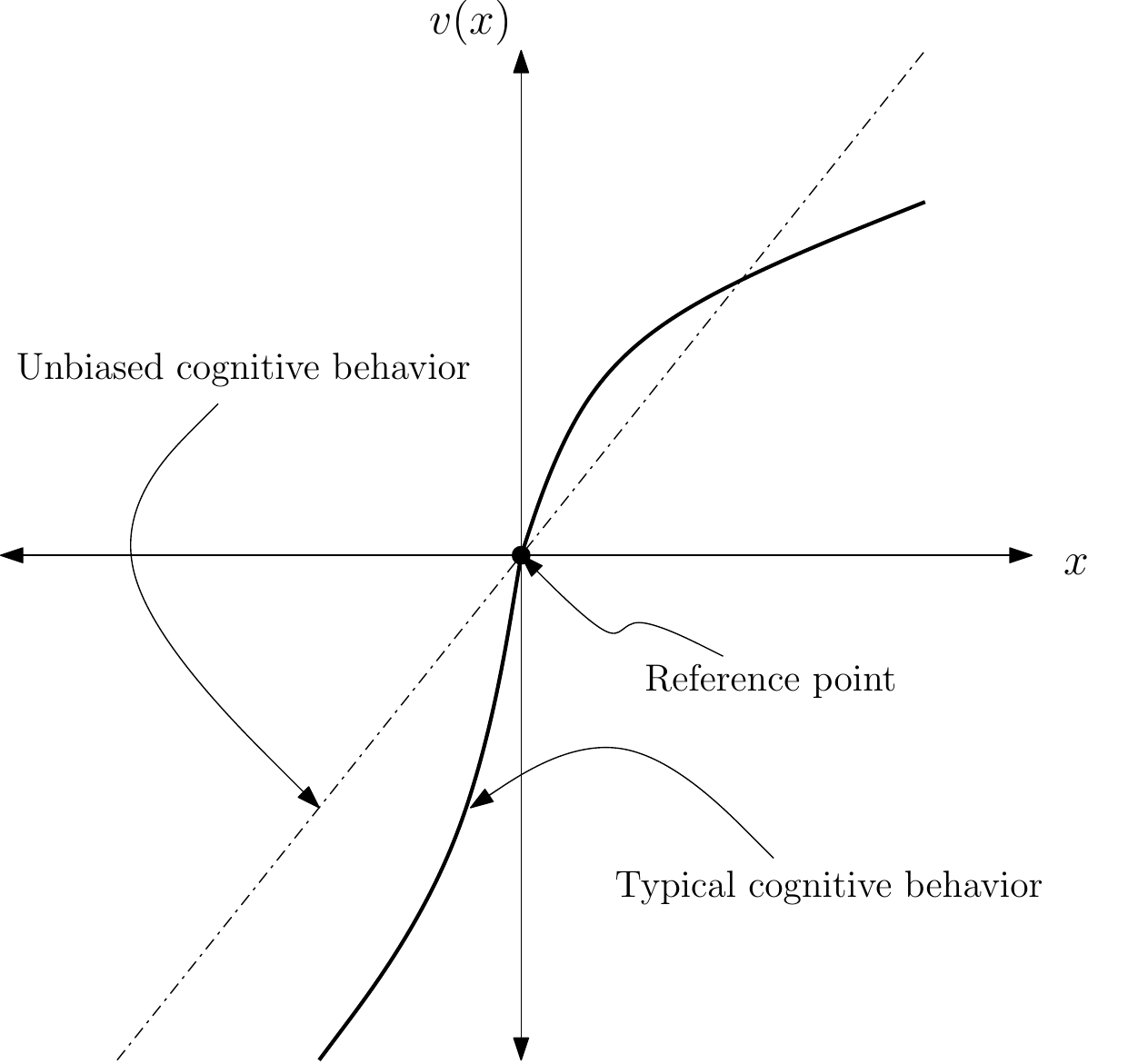}%
        \label{Fig: value}}
    }
    \caption{Typical Weight and Value Functions in Prospect Theory Models}
    \label{Fig: Prospect-Theory-Model}
\end{figure*}

Recently, several researchers have considered the problem of designing decision-theoretic systems and/or control systems with humans-in-the-loop. Some of these works model human agents using prospect theory in order to design systems in the presence of agents that portray behaviors such as endowment effect, loss aversion and anchoring, which cannot be explained by the traditional expected utility theory. In the context of \emph{decision-theoretic} systems, Nadendla \emph{et al.} have presented detection rules employed by prospect theoretic agents in \cite{Nadendla2016} under different scenarios based on decision costs. In particular, the authors have focused on two types of prospect theoretic agents, namely optimists and pessimists, and have shown that the prospect-theoretic decision rules may deviate significantly from decision rules that are optimal in the Bayesian sense. This separation between the decision rules employed by humans and unbiased agents depends on the degree of cognitive bias present in the agent's utility. In the context of \emph{communication} systems, Li and Mandayam have analyzed the behavior/impact of prospect theoretic agents in wireless networks in \cite{Li2014}, where selfish players deviate from maximizing expected utilities and employ equilibrium policies for transmission over a random access channel under throughput rewards, delay penalties and energy costs. In \cite{Yu2016}, Yu \emph{et al.} have investigated the problem of spectrum investment (whether to sense and use the spectrum without any conflict as a secondary user, or lease the spectrum from the primary user) at a prospect-theoretic user when the spectrum is available with uncertainty. In the case of \emph{power} systems, El Rahi \emph{et al.} have modeled the interaction between microgrid operators within a smart grid as a Bayesian game in \cite{ElRahi2016}, where the prospect-theoretic microgrid operator can choose to either sell its stored energy immediately, or can store for future emergencies. Sanjab \emph{et al.} have formulated in \cite{Sanjab2017} a cyber-physical security problem involving drone delivery systems and adversarial interdiction attacks, and have studied the underlying game using prospect theory which captures the subjective behaviors of the players.



\section{Preliminaries on Human Decision Making \label{sec: Preliminaries}}

The study of human decision making has gained significant interest after Von Neumann and Morgenstern first proposed a rationality model for individual human agents in \cite{Book-VonNeumann} based on maximizing the agents' respective expected utilities, and modeled interaction between multiple human agents in a game-theoretic setting. This rationality model for human decision makers based on expected utility theory has been widely accepted in solving decision problems across several disciplines. However, there have been several accounts in the behavioral economics and psychology literatures where humans significantly deviate from the aforementioned rationality model based on maximizing expected utilities. For example, behaviors such as loss aversion, framing and anchoring are not accounted for in the expected utility framework.

Some of the initial attempts to expand the horizons of expected utility theory were based on the broad class of \emph{bounded rationality} models \cite{Book-Rubinstein1998} proposed by Herbert Simon in the 1950s. These models assume that the agent's rationality can be limited broadly due to one or more reasons such as non-classical information structures, limited memory and computational complexity, to name a few. Meanwhile, there were also attempts to study human decision making using random utilities within a utility-maximization framework in order to incorporate the uncertainties regarding the knowledge of agents' utilities. Some standard examples of random-utility models include logit, nested logit, probit and generalized extreme value (GEV) models \cite{Book-Train}. 

Although several choice models have been proposed to improve expected utility theory over the following three decades, these models were not successful in comprehensively modeling human behavior, until Kahmenan and Tversky proposed an alternative descriptive model called \emph{prospect theory} for discrete choice sets in \cite{Kahneman1979}. In the following subsection, we present a brief description of Kahneman-Tversky's prospect theory model as a prelude to our analysis of the main problem in the balance of the paper.

\subsection{Prospect Theory}
According to Kahneman and Tversky in \cite{Kahneman1979}, human decision making is a distorted utility-maximization framework where the distortion in probabilities due to human cognition is modeled using a weight function and the distortion in rewards/costs due to human cognition is modeled using value functions. This can be summarized as follows.

Let an agent be presented with a set $C = \{ 1, \cdots, M \}$ of $M$ discrete choices. Let $p_i$ be the probability with which the agent decides to choose $i \in C$, and $x_i$ be the corresponding outcome for choosing $i \in C$. Then the agent experiences a utility
\begin{equation}
	V = \displaystyle \sum_{i = 1}^{M} w(p_i) v(x_i),
\end{equation}
where $w: [0,1] \rightarrow [0,1]$ is the weight function that maps the true probabilities to the agent's subjective probabilities, and $v: \mathbb{R} \rightarrow \mathbb{R}$ is the value function that maps the outcome of a given decision to its perceived value. Weight functions typically capture the human agent's introspection of the choice probabilities in a sequestered manner. Such an isolated analysis has been practically observed to result in overestimating probabilities of the rarer choices and underestimating probabilities of the more likely choices, thus resulting in a weight function with a convex-concave shape, as shown in Figure \ref{Fig: weight}. On the other hand, value functions capture the human agent's valuation of a given outcome based on its psychological introspection of gains and losses about a set reference point. This is typically observed to be an $S$-shaped function as shown in Figure \ref{Fig: value}, with a steeper slope on the losses to capture the behavior of loss-aversion. For details regarding the estimation of weight and value functions, the reader may refer to the relevant nonlinear regression methods in \cite{Gonzalez1999, Balavz2013}.

\subsection{Prospect Theory for Continuous Decision Spaces}

The downside of prospect theory model as proposed by Kahmenan and Tversky in \cite{Kahneman1979}, is that it entails discrete strategy spaces. Furthermore, very few efforts have been successful in extending these models to continuous spaces until about 10 years ago, when Rieger and Wang investigated the convergence of discretized models in continuous spaces in \cite{Rieger2008} within the context of prospect theory. In this subsection, we provide an overview of the main results in \cite{Rieger2008}, which we will use in this paper to model human decision making.

While Rieger and Wang analyzed the convergence of discretized models in continuous spaces, they had found that Schneider-Lopes' model (similar to the one proposed by Kahneman and Tversky in \cite{Kahneman1979}) does not converge. For details, the reader may refer to Theorem 2 in \cite{Rieger2008}. Interestingly, the normalized prospect-theoretic utility, as proposed by Karmarkar in \cite{Karmarkar1978}, converges to a tractable form. Therefore, in this paper, we use the model proposed by Rieger and Wang in \cite{Rieger2008}, which can be formally stated as follows:
\begin{thrm}[Adapted from Theorems 1 and 3 in \cite{Rieger2008}]
Let $p$ be a probability distribution on $\mathbb{R}$ with exponential decay at infinity, and let $p_n$ be defined for all $n \in \mathbb{N}$ as 
\begin{equation}
	p_n = \sum_{z \in \mathbb{Z}} p_{z,n} \delta_{z/n},
\end{equation}
where 
\begin{equation}
	p_{z,n} \triangleq \displaystyle \int_{\frac{z}{n}}^{\frac{z+1}{n}} dp
\end{equation}
and $\mathbb{Z}$ is the set of integers. Then, if the value function $v \in \mathcal{C}^1(\mathbb{R})$ has at most polynomial growth, and if there exists some $\alpha \in (0,1)$ and some finite number $k > 0$ such that 
\begin{equation}
	\displaystyle \lim_{\epsilon \rightarrow 0} \frac{w(\epsilon)}{\epsilon^{\alpha}} = k,
	\label{Eqn: PT-weight-alpha}
\end{equation}
then, the prospect theory utility (proposed by Karmarkar in \cite{Karmarkar1978})
\begin{equation}
	PT(p_n) = \frac{\displaystyle \sum_z w(p_{z,n}) v(z/n)}{\displaystyle \sum_z w(p_{z,n})}
\end{equation}
exists for all $n \in \mathbb{N}$ and converges to
\begin{equation}
	\displaystyle \frac{\displaystyle \int p(x)^\alpha v(x) dx}{\displaystyle \int p(x)^\alpha dx}
\end{equation}
as $n \rightarrow \infty$.

Since $p_n$ converges to $p$ in the weak * topology, one can extend $PT$ to $p$ by continuity and set
\begin{equation}
	PT(p) \triangleq \displaystyle \frac{\displaystyle \int p(x)^\alpha v(x) dx}{\displaystyle \int p(x)^\alpha dx}.
\end{equation}

\label{Thrm: PT-Continuous}
\end{thrm}

In this model, note that the parameter $\alpha$ captures the degree at which an agent overvalues low probability events. This is pointed out in Equation \eqref{Eqn: PT-weight-alpha}, where the parameter $\alpha$ is defined as a polynomial exponent that approximates the weight function effectively for very low (atomic) probability events. Other characteristics of the weight function are irrelevant in this model.

Given that the model presented in Theorem \ref{Thrm: PT-Continuous} is derived out of a well-motivated discrete choice model and has a desired form, it motivates us to consider the Rieger-Wang model presented in Theorem \ref{Thrm: PT-Continuous} in characterizing human behavior in decision theoretic frameworks.


\section{Problem Setup \label{sec: Prob-Setup}} 
\begin{figure}[!t]
	\centering
    \includegraphics[width=0.5\textwidth]{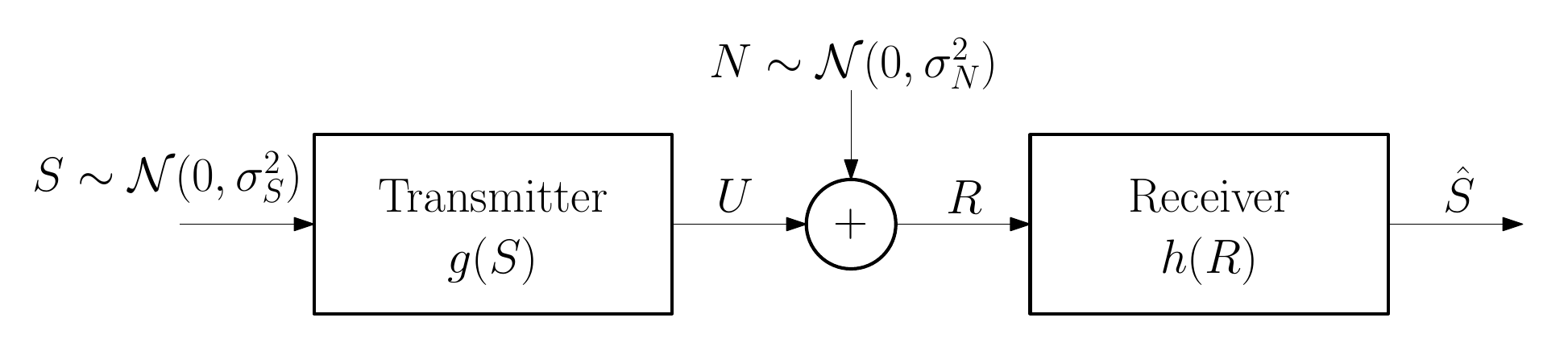}
    \caption{Strategic Communication over a Gaussian Test Channel}
    \label{Fig: model}
\end{figure}

Consider a transmitter-receiver pair, namely Alice and Bob respectively, as shown in Figure \ref{Fig: model}. Alice encodes the source signal $S \sim \mathcal{N}(0,\sigma_S^2)$ and transmits the encoded signal $U = g(S)$ to Bob over a Gaussian test channel such that her \emph{behavioral distortion}
\begin{equation}
\Scale[0.9]{
	D_T(g,h) = \displaystyle \frac{\displaystyle \iint \left[ p(S = s, R = r) \right]^{\alpha_T} v_T[S = s, \hat{S} = h(r)] \ ds \ dr}{\displaystyle \iint \left[ p(S = s, R = r) \right]^{\alpha_T} \ ds \ dr}
	\label{Eqn: Distortion-Tx} 
}
\end{equation}
is minimized. Furthermore, Alice is power-constrained, i.e., $\mathbb{E}(U^2) \leq P$. Having received a signal $R = U + N$, where $N \sim \mathcal{N}(0,\sigma_N^2)$, Bob decodes the received signal into $\hat{S} = h(R)$ so as to minimize his behavioral distortion
\begin{equation}
\Scale[0.9]{
	D_R(g,h) = \displaystyle \frac{\displaystyle \iint \left[ p(S = s, R = r) \right]^{\alpha_R} v_R[S = s, \hat{S} = h(r)] \ ds \ dr}{\displaystyle \iint \left[ p(S = s, R = r) \right]^{\alpha_R} \ ds \ dr},
	\label{Eqn: Distortion-Rx}
}
\end{equation} 
where the weighting exponents at the transmitter and the receiver are denoted by $\alpha_T$ and $\alpha_R$, respectively. For the sake of notational convenience, we sometimes ignore the type-index $k$ of the weighting parameter in $\alpha_k$ where $k \in \{ T, R \}$, and denote it as $\alpha$ whenever our arguments hold true at both the transmitter and the receiver. 

The situation above naturally defines a leader-follower game between Alice and Bob, with Alice acting as the leader. We are interested in computing a Stackelberg equilibrium for this game, i.e. a pair  $(g^*, h^*)$ such that
\begin{equation}
\begin{array}{lcl}
	h^*(g) & \triangleq & \displaystyle \argmin_{h \in \mathcal{H}} \ D_R(g,h),
	\\[2ex]
	g^* & \triangleq & \displaystyle \argmin_{g \in \mathcal{G}} \ D_T(g,h^*(g)),
\end{array}
\end{equation}
where $\mathcal{G} = \left\{ g \in \mathcal{H} \ \left| \ \mathbb{E}\left[ g(S)^2 \right] \leq P \right. \right\}$ is the set of all Borel measurable strategies that satisfy the transmitter's power constraint, and $\mathcal{H}$ is the set of all Borel measurable strategies.

As a first approach, we restrict ourselves to the space of linear policies at the transmitter, which are denoted using the parametric model
\begin{equation}
	g(S) = k_1 S + k_0,
	\label{Eqn: Tx-Linear-Policy}
\end{equation}
where $k_0 \in \mathbb{R}$ and $k_1 \in \mathbb{R}$ are Alice's strategic parameters. Furthermore, we also assume that both Alice and Bob employ squared-error as their value functions, which are given by
\begin{equation}
	v_T(S, \hat{S}) = v_R(S, \hat{S}) = ( \hat{S} - S )^2,
\end{equation}
for all $S \in \mathbb{R}$ and $\hat{S} \in \mathbb{R}$. 

\section{Equilibrium Analysis \label{sec: Analysis}}

In this section, we analyze the \emph{Stackelberg} equilibrium between Alice (leader) and Bob (follower) within the space of linear policies at the transmitter, as indicated in Equation \eqref{Eqn: Tx-Linear-Policy}. Since Alice knows that Bob attempts to minimize $D_R$, she chooses an encoding strategy (i.e. find $k_0$ and $k_1$) that minimizes $D_T$, when Bob employs its best-response decoding rule $h(R)$ that minimizes $D_R$.

Given that Alice employs a linear policy as shown in Equation \eqref{Eqn: Tx-Linear-Policy}, we have $U \sim \mathcal{N}(k_0, k_1^2 \sigma_S^2)$. Bob collects a noisy measurement $R$ of the encoded signal $U$ over an AWGN channel, and therefore, we have $R \sim \mathcal{N}(k_0, k_1^2 \sigma_S^2 + \sigma_N^2)$ and $R|S \sim \mathcal{N}(k_1 S + k_0, \sigma_N^2)$.

Therefore, we can compute the conditional probability
\begin{equation}
	\begin{array}{lcl}
		p(S = s | R = r) & = & \displaystyle \frac{p(R = r | S = s) p(S = s)}{p(R = r)}
		\\[3ex]
		& = & \displaystyle \mathcal{N} \left( \frac{k_1 \sigma_S^2 (r - k_0)}{k_1^2 \sigma_S^2 + \sigma_N^2}, \frac{\sigma_S^2 \sigma_N^2}{k_1^2 \sigma_S^2 + \sigma_N^2} \right)
	\end{array} 
\end{equation}

Given the conditional distribution $p(S = s | R = r)$, we compute the denominator of the distortion functions at both Alice and Bob in Equations \eqref{Eqn: Distortion-Tx} and \eqref{Eqn: Distortion-Rx} respectively, as shown below.

\begin{equation}
\begin{array}{l}
	\displaystyle \iint \left[ p(S = s, R = r) \right]^\alpha \ ds \ dr 
	\\[2ex] 
	\quad = \displaystyle \int \left[ p(R = r) \right]^\alpha \left( \int \left[ p(S = s | R = r) \right]^\alpha \ ds \right) \ dr 
	\\[2ex]
	\quad = \displaystyle \left( \frac{1}{\sqrt{\alpha}} \left\{ \frac{\sigma_S \sigma_N \sqrt{2 \pi}}{\sqrt{k_1^2\sigma_S^2 + \sigma_N^2}} \right\}^{1 - \alpha} \right) \int \left[ p(R = r) \right]^\alpha \ dr 
	\\[3ex]
	\Scale[0.8]{
	\quad = \displaystyle \left( \frac{1}{\sqrt{\alpha}} \left\{ \frac{\sigma_S \sigma_N \sqrt{2 \pi}}{\sqrt{k_1^2\sigma_S^2 + \sigma_N^2}} \right\}^{1 - \alpha} \right) \cdot \left( \frac{1}{\sqrt{\alpha}} \left\{ \sqrt{2 \pi \left( k_1^2 \sigma_S^2 + \sigma_N^2 \right)} \right\}^{1 - \alpha} \right)
	}
	\\[3ex]
	\quad = \displaystyle \frac{1}{\alpha} \left( 2 \pi \sigma_S \sigma_N \right)^{1 - \alpha}.
\end{array}
\label{Eqn: Denominator-Linear-Policy}
\end{equation}

Substituting Equations \eqref{Eqn: Denominator-Linear-Policy} into the distortion functions in Equations \eqref{Eqn: Distortion-Tx} and \eqref{Eqn: Distortion-Rx}, we have
\begin{subequations}
\begin{equation}
\Scale[0.8]{
	D_T(g,h) = \displaystyle \frac{\alpha_T  \displaystyle \iint \left[ p(S = s, R = r) \right]^{\alpha_T} v_T[S = s, \hat{S} = h(r)] \ ds \ dr}{\left( 2 \pi \sigma_S \sigma_N \right)^{1-\alpha_T}}
	\label{Eqn: Distortion-Tx-Linear-Policy} 
}
\end{equation}
\begin{equation}
\Scale[0.8]{
	D_R(g,h) = \displaystyle \frac{\alpha_R \displaystyle \iint \left[ p(S = s, R = r) \right]^{\alpha_R} v_R[S = s, \hat{S} = h(r)] \ ds \ dr}{\left( 2 \pi \sigma_S \sigma_N \right)^{1-\alpha_R}} 
	\label{Eqn: Distortion-Rx-Linear-Policy}
}
\end{equation}
\label{Eqn: Distortions-Linear-Policy}
\end{subequations}

Also, note that the term $\left[ p(S = s, R = r) \right]^\alpha$ inside the integral in the numerator of the distortion functions can be further simplified as follows.
\begin{equation}
	\begin{array}{l}
		\left[ p(S = s, R = r) \right]^\alpha 
		\\[2ex]
		\qquad = \displaystyle \left[ p(S = s | R = r) \right]^\alpha \left[ p(R = r) \right]^\alpha
		\\[2ex]
		\qquad = \displaystyle \displaystyle \left[ \mathcal{N} \left( \frac{k_1 \sigma_S^2 (r - k_0)}{k_1^2 \sigma_S^2 + \sigma_N^2}, \frac{\sigma_S^2 \sigma_N^2}{k_1^2 \sigma_S^2 + \sigma_N^2} \right) \right]^\alpha \cdot
		\\[3ex]
		\qquad \qquad \qquad \qquad \qquad \qquad \left[ \mathcal{N}(k_0, k_1^2 \sigma_S^2 + \sigma_N^2) \right]^\alpha
		\\[2ex]
		\qquad = \displaystyle \frac{1}{\alpha} \left[ 2 \pi \sigma_S \sigma_N \right]^{1 - \alpha} \ p_{\alpha}(S = s | R = r) \ p_{\alpha}(R = r),
	\end{array} 
\end{equation}
where
\begin{subequations}
\begin{equation}
	p_{\alpha}(S = s | R = r) = \Scale[0.9]{ \displaystyle \mathcal{N} \left( \frac{k_1 \sigma_S^2 (r - k_0)}{k_1^2 \sigma_S^2 + \sigma_N^2}, \frac{\sigma_S^2 \sigma_N^2}{\alpha \left( k_1^2 \sigma_S^2 + \sigma_N^2 \right)} \right),}
\end{equation}
and
\begin{equation}
	p_{\alpha}(R = r) = \mathcal{N} \left( k_0, \frac{k_1^2 \sigma_S^2 + \sigma_N^2}{\alpha} \right).
\end{equation}
\end{subequations}

As a result, the distortions at Alice and Bob, given in Equation \ref{Eqn: Distortions-Linear-Policy}, can be simplified into the following.
\begin{subequations}
\begin{equation}
	\begin{array}{l}
		D_T(g,h) = \displaystyle \mathbb{E}_{\alpha_T} \left\{ v_T[S,\hat{S}] \right\}
		\\[2ex]
		\Scale[0.85]{ 
		\ = \displaystyle \int p_{\alpha_T}(R = r) \int p_{\alpha_T}(S = s | R = r) v_T[S = s, \hat{S} = h(r)] \ ds dr
		}
	\end{array}
	\label{Eqn: Distortion-Tx-Linear-Policy-2} 
\end{equation}
\begin{equation}
	\begin{array}{l}
		D_R(g,h)  = \displaystyle \mathbb{E}_{\alpha_R} \left\{ v_R[S,\hat{S}] \right\}
		\\[2ex]
		\Scale[0.85]{
		\ = \displaystyle \int p_{\alpha_R}(R = r) \int p_{\alpha_R}(S = s | R = r) v_R[S = s, \hat{S} = h(r)] \ ds dr
		}
	\end{array}	 
	\label{Eqn: Distortion-Rx-Linear-Policy-2}
\end{equation}
\label{Eqn: Distortions-Linear-Policy-2}
\end{subequations}

Note that the distortions in Equation \ref{Eqn: Distortions-Linear-Policy-2} are similar in structure to the case of unbiased agents, as discussed in \cite{Akyol2017b}. In such a case, if Bob's value function $v_R(S,\hat{S}) = (S - \hat{S})^2$ is the squared error function, Bob's distortion has a very similar structure to that of MSE, averaged using the conditional distribution $p_{\alpha_R}(S | R)$. Consequently, the optimal decoding strategy at Bob is given by $\hat{S} = \mathbb{E}_{\alpha_R}(S | R)$, which is the mean of the conditional distribution $p_{\alpha_R}(S | R)$. In other words, we have
\begin{equation}
	\hat{S} = \displaystyle \frac{k_1 \sigma_S^2}{k_1^2 \sigma_S^2 + \sigma_N^2} \cdot R - \frac{k_0 k_1 \sigma_S^2}{k_1^2 \sigma_S^2 + \sigma_N^2}.
	\label{Eqn: Rx-optimal-Tx-linear}
\end{equation}

On expanding $R$ in Equation \eqref{Eqn: Rx-optimal-Tx-linear}, we have
\begin{equation}
\begin{array}{lcl}
	\hat{S} & = & \displaystyle \frac{k_1 \sigma_S^2}{k_1^2 \sigma_S^2 + \sigma_N^2} (k_1 S + k_0 + N) - \frac{k_0 k_1 \sigma_S^2}{k_1^2 \sigma_S^2 + \sigma_N^2}
	\\[3ex]
	& = & \displaystyle \frac{k_1 \sigma_S^2}{k_1^2 \sigma_S^2 + \sigma_N^2} (k_1 S + N).
\end{array}
	\label{Eqn: Rx-optimal-Tx-linear-2}
\end{equation}

Note that Bob always nullifies the impact of $k_0$. As a result, Alice finds no incentive to invest any energy into $k_0$. Therefore, without any loss in generality, we henceforth assume $k_0 = 0$. 

Since Bob uses squared-error as its value function, its best response to Alice's linear policy, as stated in Equation \eqref{Eqn: Tx-Linear-Policy}, is given by Equation \eqref{Eqn: Rx-optimal-Tx-linear}. Consequently, the distortion at Alice due to the Bob's best response strategy is given by
\begin{equation}
\begin{array}{l}
	D_T(g,h) = \displaystyle \mathbb{E}_{\alpha_T} \left[ \left( S - \hat{S} \right)^2 \right],
\end{array}
\end{equation}
where $\mathbb{E}_{\alpha_T}$ is an expectation over the distorted density function $p_{\alpha_T}(R) \cdot p_{\alpha_T}(S | R)$ due to the presence of Alice's cognitive bias $\alpha_T$. 

Note that Bob's best response strategy is given by $\hat{S} = \mathbb{E}(S|R)$. Therefore, the distortions at both Alice and Bob can be expressed in terms of the variance of $p_{\alpha_T}(s|r)$, as shown below.
\begin{equation}
\begin{array}{lcl}
	D_T(k_1) & = & \displaystyle \mathbb{E}_{\alpha_T} \left[ \left( S - \mathbb{E}(S|R) \right)^2 \right]
	\\[2ex]
	& = & \displaystyle \frac{1}{\alpha_T}  \frac{\sigma_S^2 \sigma_N^2}{ \left( k_1^2 \sigma_S^2 + \sigma_N^2 \right) } \ = \ \displaystyle \frac{\alpha_R}{\alpha_T}  D_R (k_1).
\end{array}
\end{equation}

Notice that the distortions at both Alice and Bob are monotonically decreasing with respect to $k_1$. In other words, if Alice has a power constraint, Alice always chooses $k_1 = \displaystyle \sqrt{\frac{P}{\sigma_S^2}}$ in order to minimize its distortion $D_T$.

We summarize these results in the following theorem.
\begin{thrm}
In the standard Gaussian test channel with human agents modeled using prospect theory, the following linear strategies at Alice and Bob form a Stackelberg equilibrium pair.
\begin{equation}
\begin{array}{lcl}
	U^* = \displaystyle \sqrt{\frac{P}{\sigma_S^2}} S & \mbox{and} & \hat{S}^* = \displaystyle \frac{\sigma_S\sqrt{P}}{P + \sigma_N^2} R.
\end{array}
\label{Eqn: Equilibrium-Strategy}
\end{equation}

Furthermore, the corresponding distortions at Alice and Bob in the prospect theoretic sense are respectively given by
\begin{equation}
\begin{array}{lcl}
	D_T^* = \displaystyle \frac{1}{\alpha_T}  \frac{\sigma_S^2 \sigma_N^2}{ \left( P + \sigma_N^2 \right) } & \mbox{and} & D_R^* = \displaystyle \frac{1}{\alpha_R}  \frac{\sigma_S^2 \sigma_N^2}{ \left( P + \sigma_N^2 \right) }.
\end{array}
\label{Eqn: Equilibrium-Distortion}
\end{equation}
\label{Thrm: Stackelberg-Equilibrium}
\end{thrm}

We study and discuss the implication of our results in Theorem \ref{Thrm: Stackelberg-Equilibrium} in more detail, numerically in Section \ref{sec: Discussion}.

\section{Results and Discussion \label{sec: Discussion}} 
The equilibrium strategies at both Alice and Bob stated in Theorem \ref{Thrm: Stackelberg-Equilibrium}, are the same as that of an unbiased agent. However, the cognitive biases $\alpha_T$ and $\alpha_R$ have a scaling effect on the distortions at Alice and Bob respectively. Another important point to note is that Alice does not find any leverage in the behavioral mismatch in order to improve its own utility. Given that $\alpha_T$ and $\alpha_R$ lie within the interval $[0,1]$, we find that the behavioral distortions at both Alice and Bob increase as $\alpha_T$ and $\alpha_R$ reduce to zero. In other words, an agent (either Alice, or Bob) is less satisfied than a rational agent if it has a cognitive bias $\alpha < 1$. 

\begin{figure}[!t]
	\centering
    \includegraphics[width=0.5\textwidth]{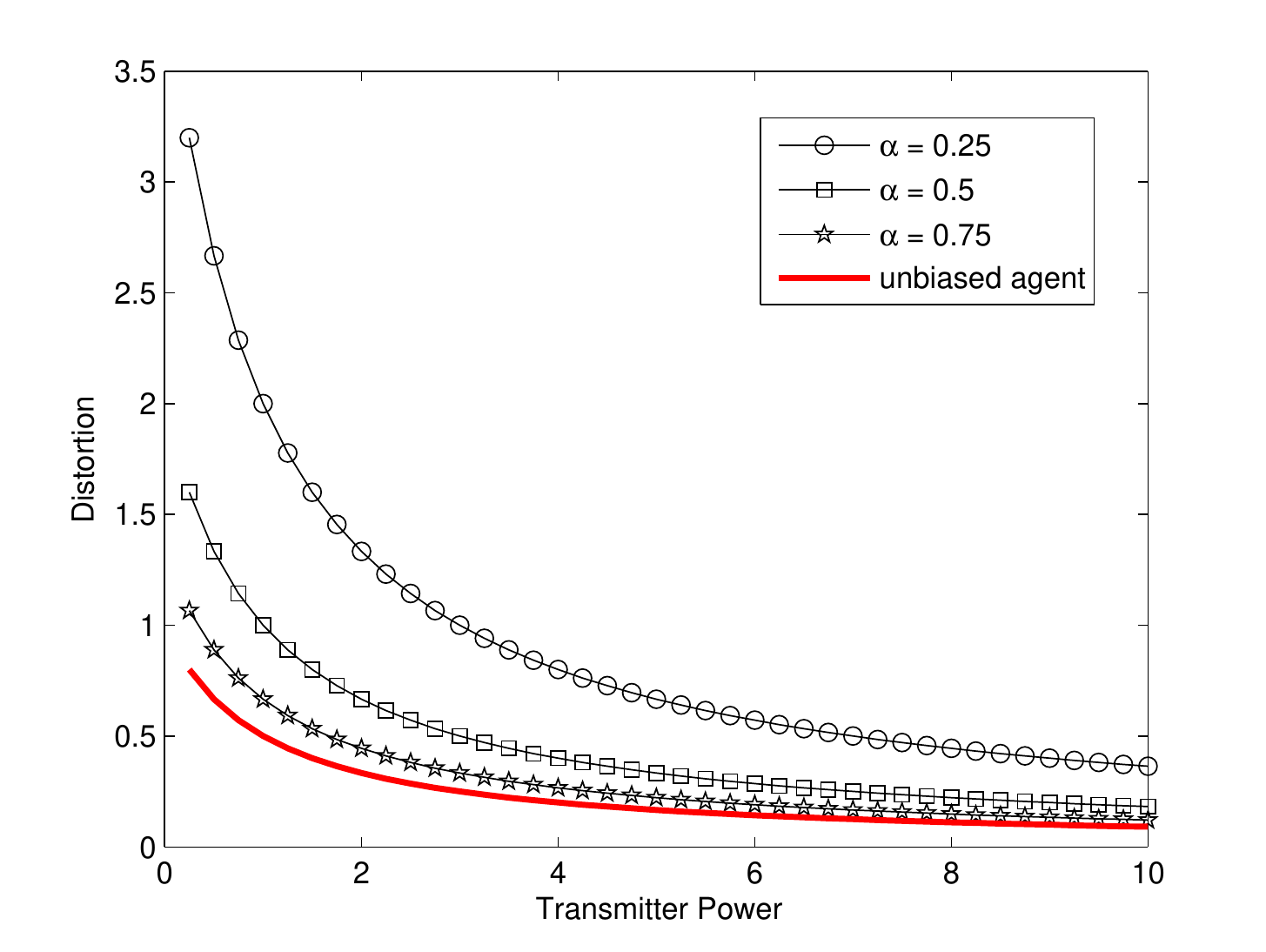}
    \caption{Distortion as a function of the transmitter power, $P$, for different values of $\alpha$}
    \label{Fig: D_vs_P}
\end{figure}

This phenomenon can be better understood from Figure \ref{Fig: D_vs_P}, where we plot the distortion function
\begin{equation}
	D = \displaystyle \frac{1}{\alpha}  \frac{\sigma_S^2 \sigma_N^2}{ \left( P + \sigma_N^2 \right) },
	\label{Eqn: Distortion-Structure}
\end{equation}
as a function of the total transmitter power $P$ for $\alpha = 0.25, 0.5, 0.75, 1$, $\sigma_S^2 = 1$ and $\sigma_N^2 = 1$. This analysis holds true for both Alice and Bob as the distortion functions in Equation \eqref{Eqn: Equilibrium-Distortion} both have the same structure as shown in Equation \eqref{Eqn: Distortion-Structure}. Note that the solid red curve in Figure \ref{Fig: D_vs_P} depicts the distortion of an unbiased agent which has $\alpha = 1$. Given that $\alpha \in [0,1]$, we observe that the distortion as perceived by an unbiased agent acts as an lower bound, and deteriorates as the value of $\alpha$ reduces to zero. In other words, although the same strategies are used by Alice and Bob respectively, regardless of $\alpha$, an agent with a smaller $\alpha$ perceive a larger distortion than the agent with a larger $\alpha$. As a result, a smaller $\alpha$ requires a larger transmitter power $P$ to achieve a prescribed level of distortion at the communicating agents. Furthermore, note that, as pictured in Figure \ref{Fig: D_vs_P}, the transmitter power $P$ has a greater impact on the agent's distortion $D$ for small values of $\alpha$. In contrast, the effect of $\alpha$ reduces with increasing transmitter power $P$, since the distortions at both Alice and Bob converge to zero asymptotically as $P \rightarrow \infty$ as depicted in our numerical results in Figure \ref{Fig: D_vs_P}. 

\section{Conclusion and Future Work \label{sec: Conclusion}}

We studied strategic communication between two human agents modeled using prospect theory. We found that behavioral agents employ the same equilibrium strategies as that of an unbiased agent, in the Stackelberg sense. We have also shown that the distortions at Alice and Bob both depend linearly on their respective weighting bias parameters. Although the transmitting agent cannot leverage the mismatch in cognitive bias with the receiver, it is the combined effect of each agent's cognitive bias and the transmitter's power budget that dictates the agent's utilities. 

In our future work, we intend to extend our results by considering a more general set of Borel strategies at the transmitter. We will also explore other channel models where the equilibrium strategies are dependent on the cognitive parameter $\alpha$. We will also investigate the effects of other cognitive biases such as value functions on the strategic interaction between Alice and Bob. Furthermore, we will also study the combined effects of both mismatched priors (beliefs) and behavioral mismatches in several interactive frameworks such as the Gaussian test channel and Bayesian deception in the presence of behavioral agents. 



\bibliographystyle{IEEEtran}
\bibliography{references}

\end{document}